\documentclass{article}
\pdfoutput=1

\AtBeginDocument{%
  \providecommand\BibTeX{{%
    \normalfont B\kern-0.5em{\scshape i\kern-0.25em b}\kern-0.8em\TeX}}}

\usepackage{graphicx}
\begin{document}

\title{Early Autism Spectrum Disorders Diagnosis Using Eye-Tracking Technology }

\author{Anna Solovyova, Sergiy Danylov, Shpenkov Oleksii, Aleksandr Kravchenko}

\begin{abstract}

While the number of children with diagnosed autism spectrum disorder (ASD) continues to rise from year to year, there is still no universal approach to autism diagnosis and treatment. A great variety of different tools and approaches for the on-site diagnostic are available right now, however, a big percent of parents have no access to them and they tend to search for the available tools and correction programs on the Internet. Lack of money, absence of qualified specialists, and low level of trust to the correction methods are the main issues that affect the in-time diagnoses of ASD and which need to be solved to get the early treatment for the little patients. Understanding the importance of this issue our team decided to investigate new methods of the online autism diagnoses and develop the algorithm that will be able to predict the chances of ASD according to the information from the gaze activity of the child. The results that we got during the experiments show supported our idea that eye-tracking technology is one of the most promising tools for the early detection of the eye-movement features that can be markers of the ASD. Moreover, we have conducted a series of experiments to ensure that our approach has a reliable result on the cheap webcam systems. Thus, this approach can be used as an additional first screening tool for the home monitoring of the early child development and ASD connected disorders monitoring. The further development of eye-tracking based autism diagnosis has a big potential of usage and can be further implemented in the daily practice for practical specialists and parents.
\end{abstract}

\providecommand{\keywords}{autism, ASD, eye-tracking, machine learning }

\maketitle

\section{Introduction and Motivation}

Autism Spectrum Disorder (ASD) is is a developmental disability characterized by persistent impairments in social interaction as well as the presence of restricted, repetitive patterns of behavior, interests, and activities of the patient. The approach to diagnose autism has changed dramatically over the past 50 years. At the beginning of 2000s, the common name autism has covered a wider range of behavioral, communication and social disorders also referred to by the umbrella term ASD, which includes autistic disorder, Asperger's syndrome and other related conditions \cite{Weintraub}. 

The number of patients with the different types of ASD symptoms continues to grow and now 1 of every 68 children in Europe is diagnosed with autism. ASD affects children from different countries and social statuses and it is reported to occur in all racial, ethnic, and socioeconomic groups. There are a lot of theories about the causes of such rocket grow of the children with diagnosed ASD, however ahead of the other theories stands the fact that in the last decade the number of the available test tools that allow to identified ASD has raised dramatically, so the probability to diagnose ASD in children also has raised. But despite the fact of the widespread usage of test tools, a lot of children remain non-diagnosed and statistical data about the percentage of the children with ASD remains biased due to the lack of available data from the developing countries. The main problems that caused such disproportion in the autism diagnosis are the absence of the practical specialists in the city, high cost of the test programs, and different difficult conditions of the children which enable the possibility to make the off-line testing. The inability to diagnose ASD at an early age is the other problem that leads to the negative consequences in the autism treatment. It is commonly known that early diagnosis and early correction programs show good results in autism treatment and can help eliminate the most severe development of ASD \cite{Koegel}. On average parents observe the first sights of autism at the age of 2 when the child establishes first social contacts with other children; however, in the majority of cases the ASD diagnoses age is more than 4 years old. It results in the situation that on average each child loses two crucial years in the ASD treatment that caused further difficulties in the child's adaptation in the social life.

It was showed that children with ASD have difficulties when needed to look directly into the partner eyes, understand the facial expression and social interactions \cite{Deborah}. For instance, in the majority of cases, the patients with ASD have trouble distinguishing fear from surprise, as well as to detect more difficult emotions\cite{Harms}. Human faces serve as an important part of the child's social adaptation and the correct perception of facial expressions is fundamental to the development of social communication \cite{Matsuda}. But the available tools for the ASD diagnosis in most cases are not able to predict how good is the child in recognizing different social situations based on understanding the core human emotions. Today as a part of the ASD symptoms treatment applied behavioral analyses are used, which can lead to good results in children when started at an early age \cite{Dawson}.

Moreover, according to the data atypical eye movements of the individuals with ASD towards various stimuli can be a good source of the information about the current state of the child development and can give good results in early ASD prediction \cite{Duan}, \cite{Nakano}. Early gaze patterns recognizing can has a big advantage for the early ASD detection and further development of the individual correction program. Moreover, the flexibility of this tool and absence of the need for additional equipment can open access for the diagnostic tools for parents all around the globe and thus increase the possibility for the success of correction programs among children of different ages.

\section{Previously works}
The inability to correctly detect human emotions and intentions is one of the most crucial aspects of the autism spectrum disorder, that leads to numerous problems in the social life of the patient. Though the individuals with ASD usually have different kinds of symptom severity and intellectual functioning, all of them have difficulties in everyday social interaction such as eye contact, engaging in reciprocal interactions, and responding to the emotional cues of other people \cite{Geraldine}. The early signs of autism such as failure to detect their own name and no interest to the other people around can be recognized during the first year of life \cite{Werner}. From the 2 to 3 additional social-oriented problems are developed - a child usually has problems with social orienting, eye contact, joint attention, imitation, responses to the emotional displays of others, and face recognition \cite{Dawson}, \cite{Mundy}. However, due to the big variety of accompanying problems, issues about the personal development path of each particular child, and limitations of the available diagnosis approaches, these problems remained undetected up to 3 or 4 years. This delay in diagnosis and treatment of autism makes the process of socialization way harder and significantly decreases the possibility of the success correction programs.

One of the most promising approaches of the autism diagnosis is the usage of eye-tracking technology to detect the deviations in the gaze behavior. Lack of eye contact stands among the most typical symptoms of ASD; it usually observed among adults and children and leads to difficulties in the emotion recognition and socialization of the person. For example, during the experiment with emotional faces it was shown that adults with ASD primarily looking 2 times less on the eye region of faces while focusing 2 times more on the mouth, body, and object regions relative to age- and verbal IQ– matched controls \cite{Klin2}. While going through the structured viewing tasks, such as looking at still faces with some kind of the particular emotions, the gaze fixation of the adults with ASD on atypical or non-feature areas of the face was also increasing ( for instance, gazing on cheeks, chin, or hairline but not at the eyes) \cite{Kevin}. In recent years a lot of the studies have been conducted to recognize the ASD children's gaze patterns and their results show that the correlation between the gaze point and the gaze fixation is also inherent to the children with diagnosed ASD. For instance, several studies have shown that children with ASD when seeing faces pay more attention to the core features of the face such as eyes and nose, in comparison to typical individuals \cite{Klin}. When observing the faces of the people with pronounced emotions children with autism tend to pay maximal attention to the eyes and mouth than other parts of the face \cite{Geest}. In his recent work A. Klin states that altered gazing patterns such as less looking at the eyes and more at mouth, body, and object areas — can serve as an indicator of the reliable quantifiers of social disability and altered engagement with the social world in autism \cite{Jones}.

It should be also mentioned that not all of the conducted to this moment studies have shown the unequivocal support to the hypotheses of the correlation between gaze behavior and ASD. During the series of experiments McPartland together with a colleague haven't found the connection between gaze fixation between adults with ASD and typically developed adults \cite{Kevin}. Thus, this area of research is yet a new branch in the ASD diagnoses that need further research to understand to what extend the eye-tracking technology can be applied to the ASD diagnosis.

\section{Experiment}

The main challenge that we faced during the experiment preparation was the necessity to make the comfort test zone for the children, which will help them concentrate on the screen and will be not boring to completely lost their willingness to take part in the experiment. Thus, to make the experiment comfortable for parents and children the screening process was held on the basis of private clinics in a cozy silent room with the soft light. To monitor the experiment the practical specialist together with a member of our team was present to ensure that the screening process was conducted correctly. Children aged from 4-6 were accompanied at all times by a parent or primary caregiver. To begin the experimental session, a participant and caregiver entered the laboratory room while a member of our team set up the application and eye-tracker. The child was buckled into a comfortable seat mounted on a pneumatic lift so that viewing height (line of sight) was standardized due to the determined level of the participant's eyes. The lights in the room were muted so that the child will be able to concentrate only on the computer monitor. During the experiment session, the experimenter (a member of our team) was concealed from the child’s view but was able to monitor the conduction of the experiment as well as the gaze patterns of the child (Figure 1).

\begin{figure} [h]
    \centering
    \includegraphics [scale=0.3]{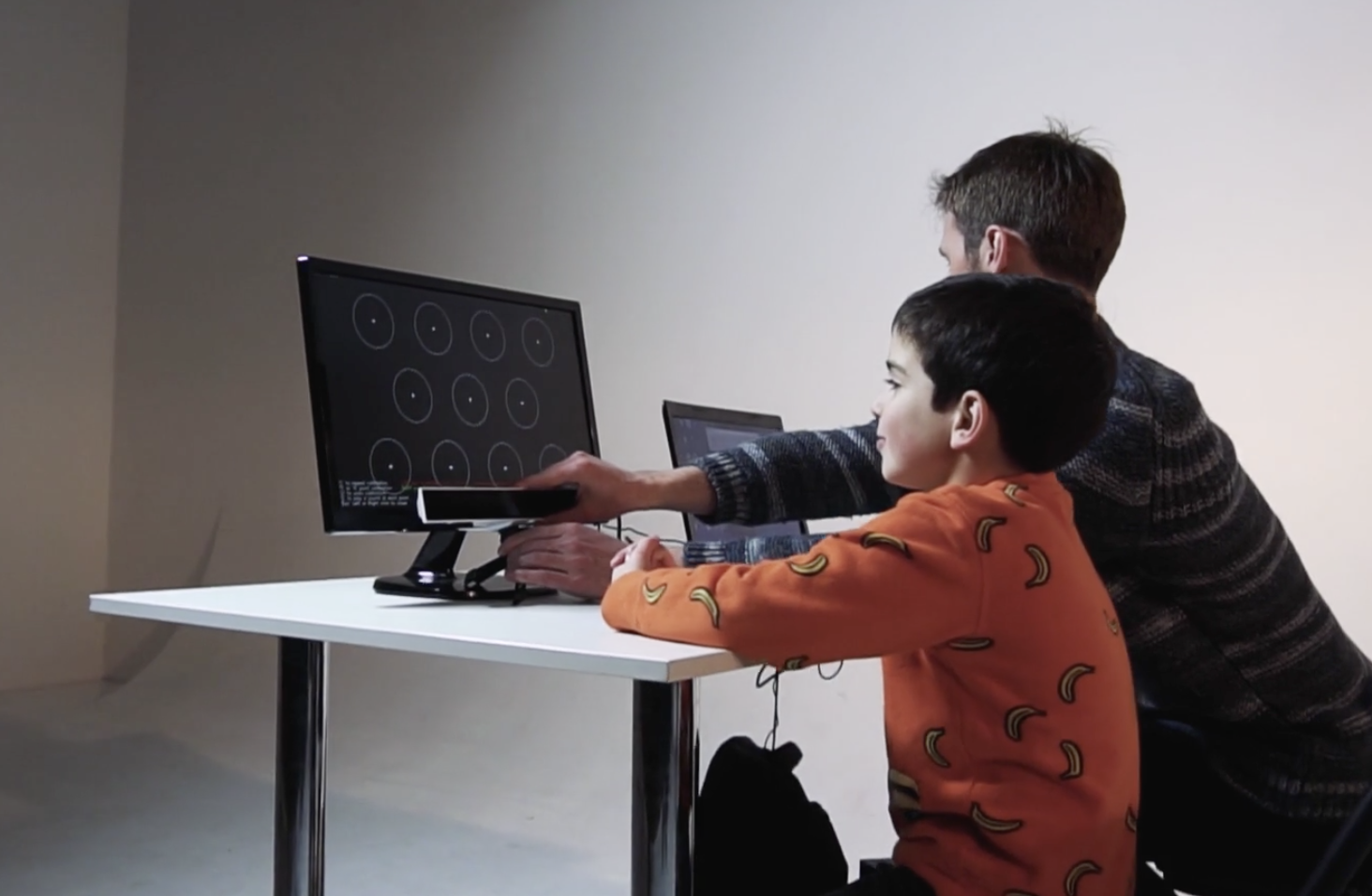}
    \caption{During the setup for the experiment}
    \label{fig:my_label6}
\end{figure}

Gaze fixation patterns were measured using eye-tracking equipment, described in the “Data Acquisition and Analyse” section. At the beginning of the data collection process, the child was comfortably seated in the chair watching a few special videos for the eye-tracker calibration. This video contains different shapes and figures of black and white color which appeared randomly on the screen. While testing the children with ASD the average amount of the additional calibration procedure was about 5 times (in the case of the high activity of the child during the examination session). To make the examination more relaxed for the children with ASD in most cases they were seated together with the parent or a care-giver. When testing the children with typical development(TD) no additional calibration was used and they were seated alone on the chair. 

During the screening session, the children observed different kinds of human faces (both male and female) with different kinds of emotions (here we should add the source of these images and why exactly they were used for our experiment). Besides the actor faces the photo of the family members of each particular child was also added to the video, so we were able to estimate the difference between the perception of the unknown faces vs faces of the caregiver, whom the child sees every day. At the end of the experiment practical specialist together with members of our team made a short review for each of the parents to help them better understand the results of their child and gain their feedback about the procedure.

\section {Participants}
  
Taking into account that the aim of our experiment was to gain the dataset of the gaze points of the children with ASD in compare to the gaze behavior patterns of the typically developed children, all the participants of the experiment were divided into two groups: children with ASD (earlier diagnosed by the practical specialist) and typically developed children. The age range of the participants was from 3 to 10 years old.

To enlarge the local community of the ASD children and to make our dataset as wide as possible, we decided to post the invitation for the screening among the dedicated to the children's ASD groups on the Facebook social platform. We have received about 150 requests for the experiment participation, but as we have limitations of the eye-tracking due to the severity of the ASD symptoms, we were able to choose only 50 participants with ASD among them. 

Typically developed children were screened on the basis of the private school "Afina" according to the previous agreement with the management of the school. The experiment session was complying with all specified screening requirements (described in the previous section). Before the start of the experiment, all parents of the children that took part in the screening were aware of the procedure and the goals of the experiment. The age groups for the TD children were the same as for the children with ASD to get the most comparable data.

\section {Data Acquisition and Analyses}

Over the past 10 years, due to the appearance of high-quality cameras in laptops and smartphones, eye-tracking has become possible on built-in cameras. However, it has limitations in accuracy. In particular, for this experiment, the error could be up to 5per cent in both x and y coordinates, for a standard webcam that shoots at 720p. Since the data was collected on a professional eye tracker, Gaussian noise was added to simulate the data from the built-in camera. Time distributions by zones on the data with and without the noise you can see in Figure 2.

\begin{figure} [h]
    \centering
    \includegraphics [scale=0.5]{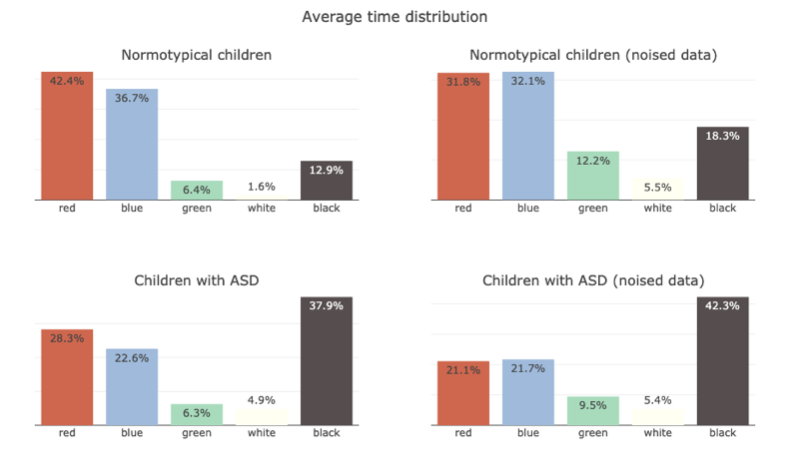}
    \caption{Average time distribution per zones}
    \label{fig:my_label}
\end{figure}

As we can see the bars of the diagram for the noised data approached the mean values, but the overall picture has not changed. Thus, it became interesting to check whether it is possible to train the algorithm on noisy data and what accuracy it will show. However, we cat not use fixations as input features, since we did not have them for noisy data. We could have merged the generated data into fixations manually, but since the data is artificial, for the correctly selected hyperparameters, we would get exactly the same fixations as for raw data.
To test the hypothesis, we used a simple fully connected network. As input data, we use time distribution by zones for each respondent. Since the dataset was very small and has only 5 features, it makes no sense to use the training results as a ready-made algorithm, but it was interesting to analyze the learning process of the neural network. You can see the classification error and quality for clean and noisy data in Figure 3 (a, b)

\begin{figure} [h]
    \centering
    \includegraphics [scale=0.3]{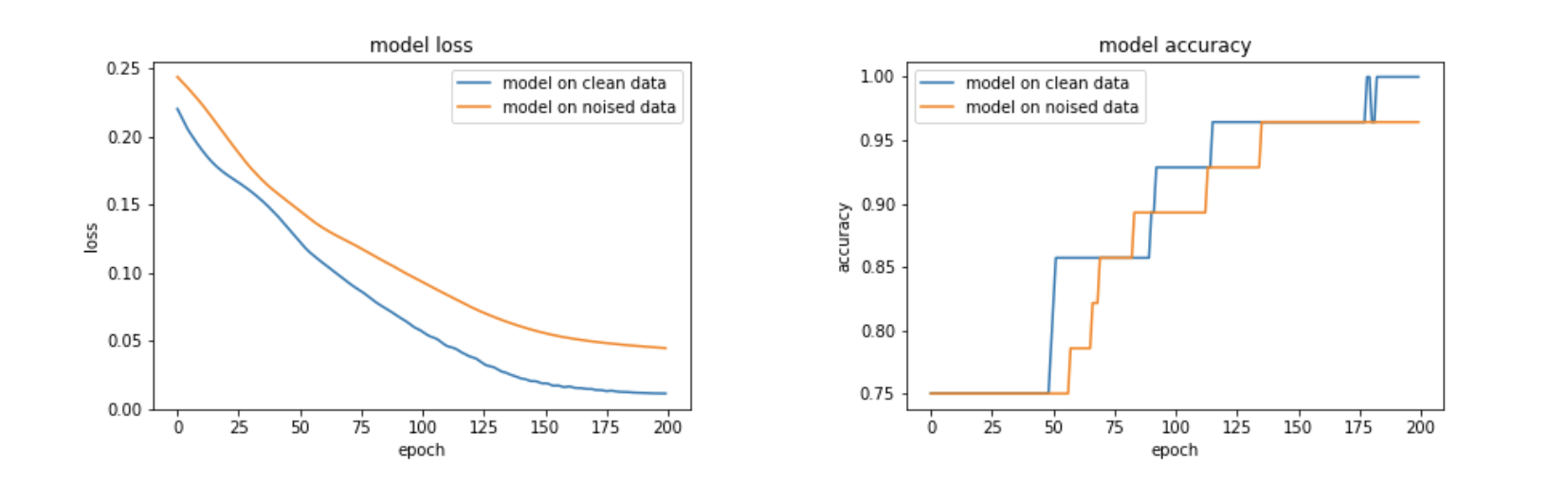}
    \caption{ Model loss and model accuracy on clean and noised 
data per training epochs}
    \label{fig:my_label1}
\end{figure}

According to Figure 3, we see that the learning process for noised data is slower and there is no way to achieve 100per cent accuracy. However, it can be seen that even on data with noise, we can classify children with sufficiently high accuracy with only 5 features. Thus, it makes sense to conduct a similar experiment on a conventional built-in camera and build an algorithm for more features.

During the experiment, we have observed marked differences in the gaze behavior of the children with ASD in comparison to the typically developed children. The core difference lay in the duration of the visual fixation on the face region of the observed actor (actress). It is known that that looking at a person's face is accompanied by a specific pattern of gaze movement \cite{Fumihiro}. The most crucial point for face scanning and thus emotion recognition such as eyes and mouth of the person were our primary focus of observation. Any deviation from the standard gaze pattern can be a sign of the disability to understand and evaluate the context of the situation properly. 

As you can see in Fig.2 and Fig.3 children with diagnosed ASD tend to pay more attention to the objects outside of the face zone, while gaze points on the mouth and eyes are very short. 

\begin{figure} [h]
    \centering
    \includegraphics [scale=0.1]{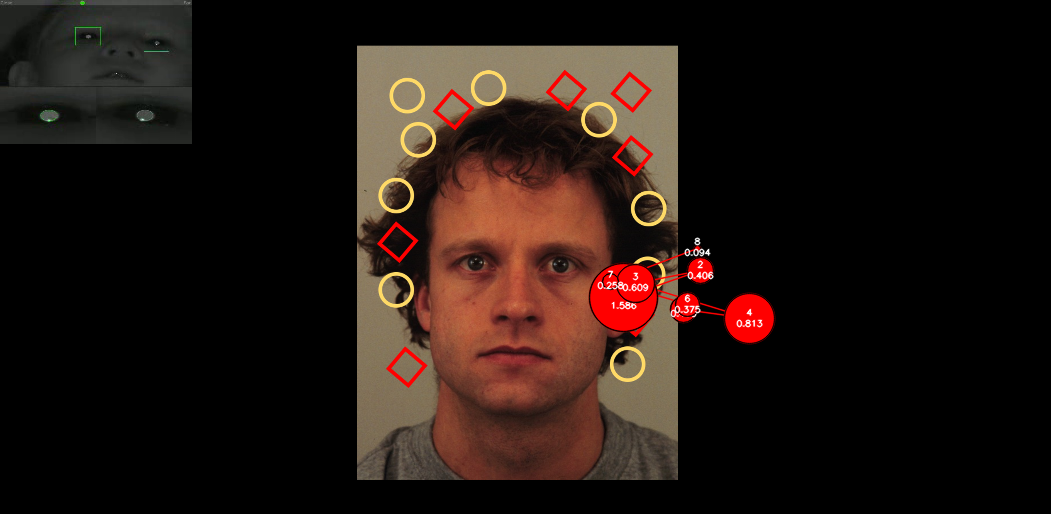}
    \caption{Example of the gaze fixation on the unknown face (ASD child)}
    \label{fig:my_label2}
\end{figure}

 \begin{figure} [h]
    \centering
    \includegraphics [scale=0.1]{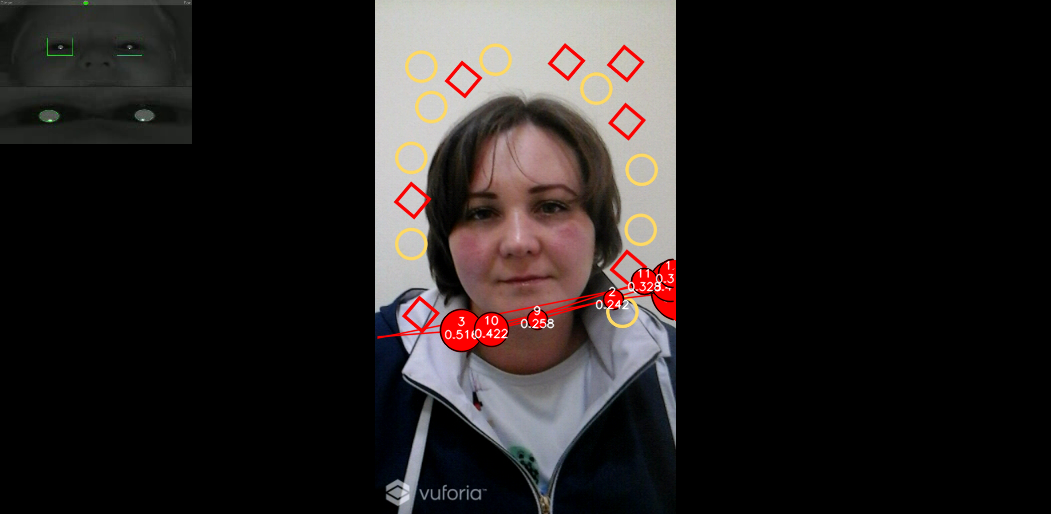}
    \caption{Example of the gaze fixation on the known face (ASD child))}
    \label{fig:my_label3}
\end{figure}

 Children with typical development look more often on the face of the person rather than on the objects around the face and the time of the gaze fixations shows the interest of the children on the mouth and eyes of the dialogue partner (Fig. 4 and Fig. 5). This fact indicates that typically developed children can easily understand the emotions of the person and the contest of the situation, using the face of the person as the representation of the inner intentions. However, children with ASD tend to miss up the face signals of the dialogue partner, which leads to the difficulties of emotion recognition and understanding. 
 
\begin{figure} [h]
    \centering
    \includegraphics [scale=0.1]{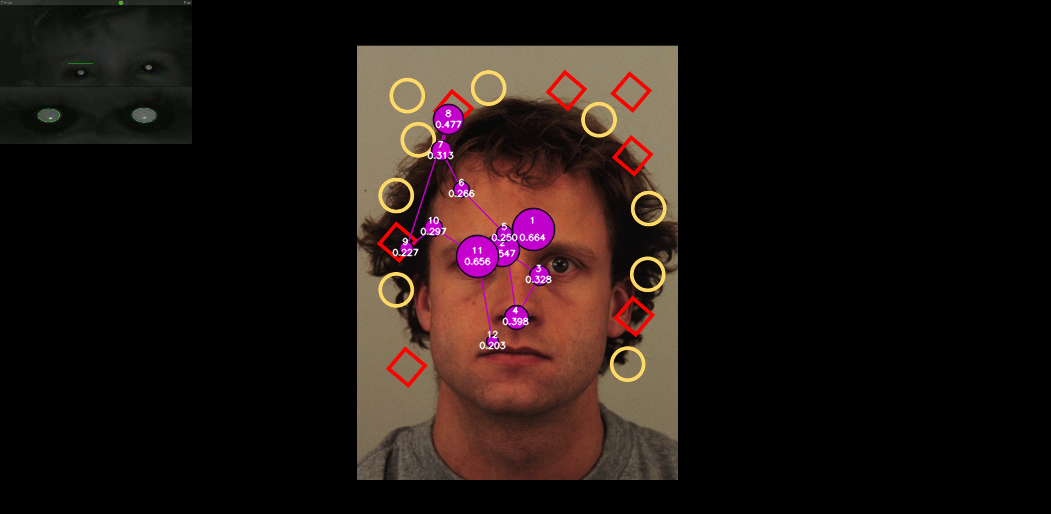}
    \caption{Example of the gaze fixation on the unknown face (TD child)}
    \label{fig:my_label4}
\end{figure}

 \begin{figure} [h]
   \centering
    \includegraphics [scale=0.1]{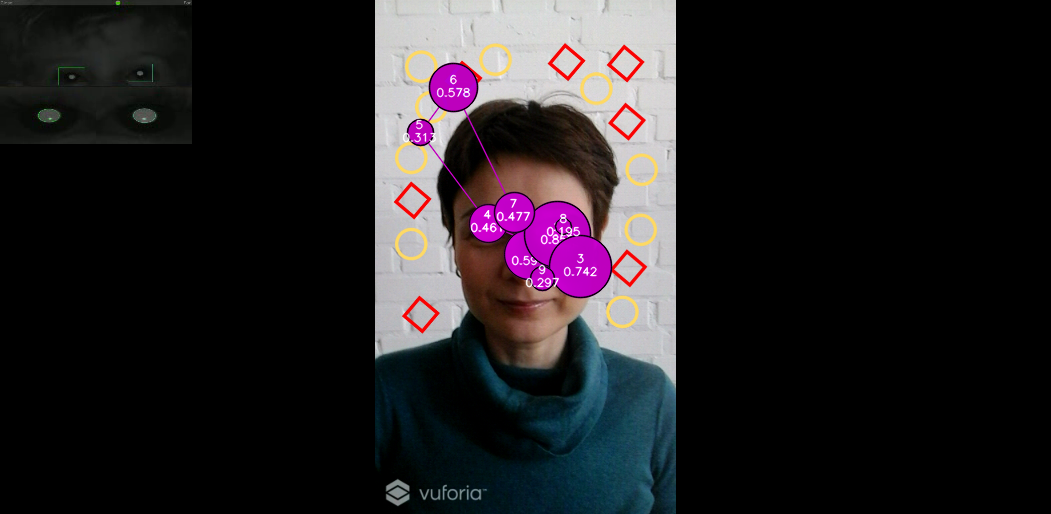}
    \caption{Example of the gaze fixation on the known face (TD child))}
    \label{fig:my_label5}
\end{figure}

\section{Conclusion}

Understanding the gaze behavior of the children and measuring the deviations in the individual gaze pattern of the child can be a good source of the information for the clinical diagnosis of ASD. The data that we have got during the experiment sessions prove the theory that children with autism spectrum disorders tend to avoid direct gazing to the face of the dialogue partner and tend to spend more time observing the objects on the periphery. The most important source of the social and emotional information of the person such as eyes and mouth remain unobserved by the ASD children that lead to an incorrect understanding of the socio-emotional context of the particular situation and as a result of the difficulties in the day-to-day interactions with other people. It was also demonstrated that children with ASD have the similar problems when looking to the faces of the family members (in our experiment it was moms and nannies), so the communications within the family can be also difficult due to the limited understanding of the emotions and feelings of the family members.

The main problem that we faced during the first stage of the experiment is the limitation of the available hardware. The eye-tracking systems as the ones that were used in our experiment (add the description of the eye-tracker) are mainly used in the research laboratories, but they almost nit available for the family home-using. In-build video cameras that are used in phones and notebooks in most cases are not useful for tiny gaze movements detection and to use them the user needs to make eye-calibration a few times during the session. This limits the usage of the algorithm in the non-laboratory environment, but new cameras that are used in the new generations of smartphones can be good support for this ASD detection approach in the future.

It should also be mentioned that eye-tracking methodology can be used as a kind of tool for the individual correction programs for children with autism. Thus, a lot of the application available today give a big variety of the emotion recognition task, they don't track the gaze activity of the child during the working sessions, so it gives almost no data about the developing of the gaze behavior of the child with ASD during the simulation sessions.

However, to this date the proposed algorithm of the ASD diagnosis cant be used as a separate tool for the clinical results, it can be a good screening tool for the early understanding of the possible autism spectrum disorders cases. Creating the basis for further clinical observations can be a good supportive option for the practical specialist. To achieve better more precise results in the ASD prognoses our team plans to gather more experiment data during the next stage of the research and use some modifications in the video screening scenario that will allow us to use a wider range of emotions and simulate real-life situations.

\bibliographystyle{unsrt}       
\bibliography{literature} 

\end{document}